\documentclass[numberedappendix,twocolumn,twocolappendix,apj]{openjournal}

\pdfoutput=1

\usepackage[utf8]{inputenc}
\usepackage{graphicx}
\usepackage{hyperref}
\usepackage{comment}
\usepackage{amsmath}
\usepackage{ amssymb }
\usepackage{lipsum} 
\usepackage{float}
\usepackage{cleveref}
\usepackage[caption=false]{subfig}
\usepackage{placeins}
\usepackage{xcolor}
\usepackage{subfig}
\usepackage[normalem]{ulem}

\newcommand{\sfourwide}{S4-Wide}
\newcommand{\sfourdeep}{S4-Ultra deep}

\def\thethreehundred{{\sc The Three Hundred}}
\def\GIZ{\textsc{Gizmo-SIMBA}}
\def\GM{\textsc{Gadget-MUSIC}}
\def\GX{\textsc{Gadget-X}}

\begin{document}

\title{Prospects for studying the mass and gas in protoclusters with future CMB observations}
\shorttitle{Protoclusters in the CMB}

\author{Anna Gardner$^{1}$}
\author{Eric Baxter$^{1\star}$}
\author{Srinivasan Raghunathan$^{2}$}
\author{Weiguang Cui$^{3}$}
\author{Daniel Ceverino$^{4,5}$}

\affiliation{$^{1}$Institute for Astronomy, University of Hawai`i, 2680 Woodlawn Drive, Honolulu, HI 96822, USA}
\affiliation{$^{2}$Center for AstroPhysical Surveys, National Center for Supercomputing Applications, Urbana, IL 61801, USA}
\affiliation{$^{3}$Institute for Astronomy, University of Edinburgh, Royal Observatory, Edinburgh EH9 3HJ, United Kingdom}
\affiliation{$^{4}$Universidad Autonoma de Madrid, Ciudad Universitaria de Cantoblanco, E-28049 Madrid, Spain}
\affiliation{$^{5}$CIAFF, Facultad de Ciencias, Universidad Autonoma de Madrid, E-28049 Madrid, Spain}

\thanks{$^{\star}$e-mail:ebax@hawaii.edu}

\shortauthors{Gardner et al.}

\begin{abstract}
Protoclusters are the progenitors  of massive galaxy clusters.  Understanding the properties of these structures is important for building a complete picture of cluster formation and for understanding the impact of environment on galaxy evolution.  Future cosmic microwave background (CMB) surveys may provide insight into the properties of protoclusters via observations of the thermal Sunyaev Zel'dovich (SZ) effect and gravitational lensing.  Using realistic hydrodynamical simulations of protoclusters from the \thethreehundred\ Project, we forecast the ability of CMB Stage 4-like (CMB-S4) experiments  to detect and characterize protoclusters with observations of these two signals.  For protoclusters that are the progenitors of clusters at $z = 0$ with $M_{200c} \gtrsim 10^{15}\,M_{\odot}$ we find that the \sfourdeep\ survey has a roughly 20\% chance of detecting the main halos in these structures with ${\rm SNR} > 5$ at $z \sim 2$ and a 10\% chance of detecting them at $z \sim 2.5$, where these probabilities include the impacts of noise, CMB foregrounds, and the different possible evolutionary histories of the structures.  On the other hand, if protoclusters can be identified using alternative means, such as via galaxy surveys like LSST and \textit{Euclid}, CMB-S4 will be able to obtain high signal-to-noise measurements of their stacked lensing and SZ signals, providing a way to measure their average mass and gas content.  With a sample of 2700 protoclusters at $z = 3$, the CMB-S4 wide survey can measure the stacked SZ signal with a signal-to-noise of 7.2, and the stacked lensing signal with a signal-to-noise of 5.7.  Future CMB surveys thus offer exciting prospects for understanding the properties of protoclusters.
\end{abstract}  


\maketitle

\vspace{1cm}

\twocolumngrid

\section{Introduction}\label{sec:intro}

Galaxy clusters are the most massive gravitationally bound structures in the Universe, with masses of roughly $10^{14}$ to $10^{15}\,M_{\odot}$ and virial radii of order 1~Mpc.  Although the definition of a \textit{protocluster} is not entirely consistent across the literature, one common and practical definition is that a protocluster is a structure that will collapse into a galaxy cluster by $z \geq 0$ \citep[see also][]{Muldrew:2015}.   By studying protoclusters, we can learn about the evolutionary histories of galaxy clusters, explore the impact of environment on galaxy formation \citep[e.g.][]{Steidel:2005,Muldrew:2017}, learn about processes driving reionization \citep[e.g.][]{Chiang_2017}, and more \citep[for a review see][]{Overzier:2016}.

Protoclusters are difficult to detect because they are at high redshift, typically reside in halos that are less massive than virialized clusters at low redshift, and they can extend over tens of comoving Mpc.  Protoclusters are also rare, having formed from extreme density fluctuations.  Detecting protoclusters therefore  requires wide-field surveys in addition to high sensitivity.
For these reasons, the number of protoclusters that have been detected to date is small, on the order of 100 (and of order 10 above $z\sim 4$).  Several techniques have been used to detect protoclusters, including looking for overdensities of galaxies selected via e.g. dropout methods in photometric surveys  \citep[e.g.][]{Toshikawa:2018}, looking for overdensities of star-forming galaxies at high redshift via their redshifted emission \citep[][for example]{Clements:2014, miller18},
and  looking for correlated Lyman-$\alpha$ absorption along lines of sight to background galaxies \citep{Newman_2022}. 

The majority of protoclusters found to date have been identified via galaxy overdensities.  However, protocluster selection via galaxy overdensities depends strongly on the galaxy sample, and the identified overdensities may or may not collapse into massive clusters at $z=0$ \citep{Cui2020}.  Moreover, the completeness and purity of protocluster samples identified via galaxy overdensities depend strongly on e.g. the choice of aperture used to identify the overdensities \citep{Muldrew:2015, Lovell:2018}.  The purity and completeness of protocluster regions identified via galaxy density has also been investigated using The300 simulations in \citet{Cui2020}.  The galaxy populations within protoclusters are  often used to form rough mass estimates of these structures by, for instance,  using stellar mass--halo mass relations \citep[e.g.][]{Laporte:2022} or by relating the galaxy overdensity to the underlying mass overdensity by assuming a value for the galaxy bias \citep[e.g][]{Toshikawa:2012}.  These estimates typically carry large uncertainties and often rely on assumptions about the member galaxy properties.

\begin{figure}
    \centering
    \includegraphics[scale= 0.35]{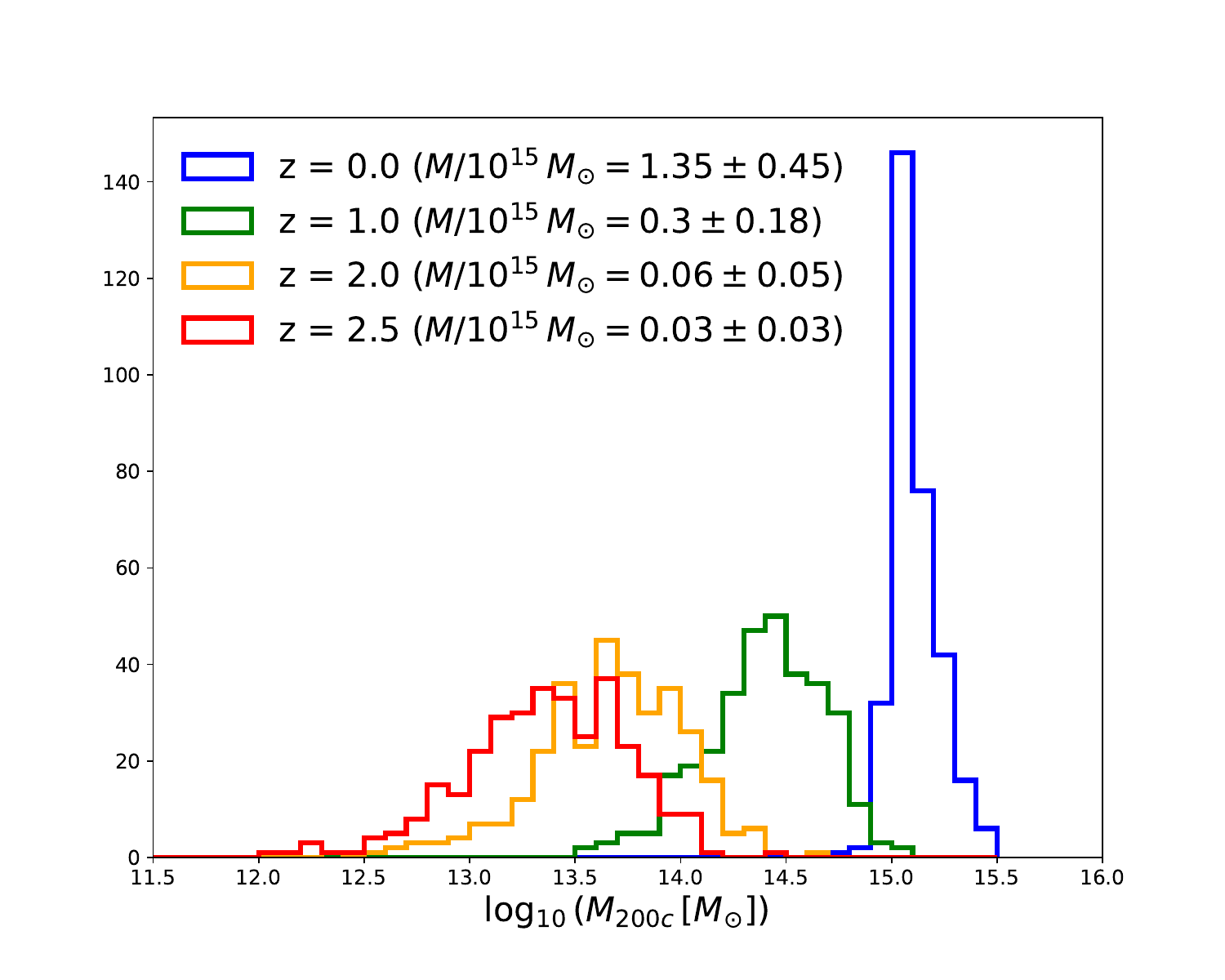}
    \caption{The redshift evolution of the mass distribution of the main progenitor halos of clusters in The300 simulations, which are selected based on a mass cut of $M \gtrsim 10^{15} M_{\odot}$ at $z = 0$.  At high redshift, the main halo in a  protocluster may constitute only a small fraction of the structure's total mass.   The numbers in parentheses in the legend indicate the mean and standard deviation of $M_{200c}$ for each redshift, in units of $10^{15}\,M_{\odot}$.
    \label{fig:mass_distribution}
    }
\end{figure}

The aim of this work is to investigate the prospects for detecting and characterizing the properties of protoclusters \textit{without} using their member galaxies.  Instead, we consider how the cosmic microwave background (CMB) can be used to study protoclusters.  Because the CMB originates from very high redshift ($z \sim 1089$), it provides a backlight to all clusters and protoclusters in the Universe.  CMB photons travelling near to clusters and protoclusters can be perturbed by these structures in several different ways.  We focus on two such effects here.  First, CMB photons can inverse Compton scatter with populations of thermal electrons in protoclusters, leading to the thermal Sunyaev Zel'dovich (SZ) effect \citep{Sunyaev:1972}.  The amplitude of this effect is proportional to the Compton-$y$ parameter, which is sensitive to the line-of-sight integral of the electron gas pressure.  The SZ effect is routinely used to detect galaxy clusters with current CMB data, providing catalogs of clusters with masses $M \gtrsim {\rm few} \times 10^{14}\,M_{\odot}$ out to redshifts of $z \sim 1.5$ \citep[e.g.][]{Bleem:2015,Hilton:2021,Planck:clusters}.    Since protoclusters are expected to have significant ionized gas content, they may also produce detectable SZ signatures.  However, detecting the SZ signals from protoclusters will be challenging for several reasons.  For one, the amount of mass in a protocluster that is contained in a single virialized halo is significantly less than the  mass of the low-redshift cluster into which the structure will evolve \citep{Muldrew:2015}.  As a result, the SZ signals from  protoclusters are expected to be much lower than those of low-redshift clusters.  The expected scaling of the SZ signal, $Y$, with halo mass, $M$, is $Y \sim M^{5/3}$ for self-similar halos \citep{Kaiser:1986} (see e.g. \citealt{Pratt:2020} for recent observational confirmation of this prediction).  Since the typical cluster with $M \sim 10^{15}\,M_{\odot}$ at $z = 0$ grew from a protocluster with a main halo of mass $\sim 10^{13.6}$ at $z = 2$ (see Fig.~\ref{fig:mass_distribution}), the average protocluster at $z = 2$ will have an SZ signal which is roughly a factor of 200 lower than that of a typical $z = 0$ cluster.  Since typical wide-field SZ surveys like those from the South Pole Telescope (SPT; \citealt{Bleem:2015}), the Atacama Cosmology Telescope (ACT; \citealt{Hilton:2021}) and {\it Planck} (\citealt{Planckszclusters}) detect clusters with signal-to-noise of order tens, it is not too surprising that detections of many protoclusters have not been reported by these surveys.  This calculation ignores mass outside of the main halo, but doing so is reasonable on account of the $M^{5/3}$ scaling of the SZ signal. Since protocluster halos may not be self-similar, this scaling is also not expected to hold exactly, but should give a reasonable indication of the challenge associated with high-redshift protocluster detection via the SZ \citep[see][for a prediction of steeper slope]{Sembolini2014}. Moreover, we expect halos at high redshift to be less concentrated than their low-redshift counterparts, leading to additional suppression of the SZ signal.\footnote{Halo concentration is positively correlated with SZ signal at fixed halo mass, see e.g. \citet{Baxter:2023}.} Despite these challenges, detecting the SZ signals from protoclusters is an exciting prospect, as such a detection would provide a  window into the gas content and thermal properties of these structures.  Indeed, during the completion of this work, the detection of the SZ signal from a protocluster was reported by \citet{Mascolo:2023}.

A second way that clusters and protoclusters may leave an imprint on the CMB is via gravitational lensing.  The paths of CMB photons travelling near massive objects like clusters or protoclusters will be perturbed, leading to an observable impact on the CMB \citep{Seljak:2000,Dodelson:2004,Lewis:2006,hu07}. 
 Gravitational lensing of the CMB by galaxy clusters can now be detected and used to calibrate cluster mass-observable relations \citep{madhavacheril15, Baxter:2015,Baxter:2018, raghunathan19, raghunathan19c, madhavacheril20}.  We consider here the possibility of using CMB lensing to constrain the masses of protoclusters.  Gravitational lensing in general is a powerful tool for measuring the masses of (proto)clusters because it is sensitive to all forms of mass, including dark matter.  Mass estimation methods that rely on visible matter, on the other hand, must make assumptions about the connection between the visible matter and the underlying dark matter, which makes up most of the mass in these structures (e.g. that the visible and dark matter are in virial equilibrium).  Such assumptions are likely to be particularly bad for protoclusters, which may still be in the process of forming, and which may contain galaxies that differ significantly from those at low redshifts or in the field.  CMB lensing has the potential to be a particularly powerful  tool for measuring protocluster masses because  --- unlike galaxy lensing --- it can be measured around structures at very high redshift.  Gravitational lensing measurements relying on galaxy lensing, on the other hand, will be challenging to apply to protoclusters because they will require measurements of the shapes of background galaxies at higher redshifts than the protoclusters \citep{Madhavacheril:2017}.

In this work, we use realistic hydrodynamical simulations of protoclusters from \thethreehundred\ Project\footnote{\url{http://the300-project.org/}} \citep[The300 for short;][]{Cui2018} to investigate the extent to which future CMB surveys can measure the SZ and CMB lensing signals from protoclusters.  Our use of hydrodynamical simulations to make these forecasts is novel and important because actively forming protoclusters may not be described by simple analytical models.  Previous work making similar forecasts \citep[e.g][]{raghunathan22} has generally relied on analytical models developed for virialized clusters in order to make forecasts for protocluster measurements.  Moreover, with these simulations, we can explicitly investigate the impact of each protocluster's evolutionary history on its SZ and lensing signals.  We will find that there is significant scatter in these signals as a result of different evolutionary histories, an effect which is often missed in analytic forecasts for protocluster measurements. While some details of protocluster evolution (such as the impact of AGN feedback and environment-dependent quenching of star formation) remain poorly understood and thus unlikely to be fully captured in the simulations, we expect that protocluster mass distributions and bulk gas properties will be accurately modeled.  It is these quantities which are relevant to determining the SZ and CMB lensing signals.  In any case, our intent here is to make a forecast for the expected observational signatures of protoclusters given a specific model (i.e. the \GX\ model in The300 simulations); departures from this model in real protoclusters  would  be interesting.

Near-term CMB surveys including Simons Observatory \citep{SimonsObservatory} and CMB Stage 4 \citep{CMBS4} will map large areas of the microwave sky to unprecedented depth.   We will show that these observations can place interesting constraints on the SZ and CMB lensing signatures of protoclusters.  We consider two possibilities: (1) detecting protoclusters with the CMB survey and characterizing their individual properties, and (2) a ``stacking'' approach, where some other survey --- such as a galaxy survey --- is used to identify protocluster locations, and the CMB survey is then used to constrain the average mass and gas properties of the resultant samples.

As mentioned above, CMB surveys like {\it Planck} have already been used to detect protoclusters by searching for emission from protocluster member galaxies \citep{Clements:2014}.  These detections rely on the high-frequency channels of {\it Planck} (353 to 857 GHz) to detect infrared emission from dusty, star-forming galaxies that has been highly redshifted.  Here, in contrast, we focus on detecting signals from the impact of protoclusters on CMB photons, rather than emission directly from the protoclusters themselves.  These ``indirect'' signals have the advantage that they are largely independent of assumptions about the protocluster member galaxies.

The organization of the paper is as follows: In \S\ref{sec:simulations} we discuss the simulation data products that are used in this work and our analysis of these data products, in \S\ref{sec:results} we present our main results, and we conclude in \S\ref{sec:discussion}.  Throughout this work we adopt a flat $\Lambda$ and cold dark matter cosmological model with parameters consistent with those of The300 \citep{Cui2018}, which are in turn consistent with the results of the \textit{Planck} mission: $h = 0.678$, $\Omega_{\rm m} = 0.307$, $\Omega_{\rm b} = 0.048$, $n_s = 0.96$, $\sigma_8 = 0.8228$ \citep{planck20_cosmo}. Unless specified otherwise, halo mass refers to $M_{200c}$, i.e. the mass contained within a sphere of radius $R_{200c}$ centered on the halo, where $R_{200c}$ is the radius at which the mean enclosed density is 200 times the critical density of the Universe at the redshift of the halo.

\section{Simulated protocluster observations}
\label{sec:simulations}

\subsection{The300 project}

The300 Project \citep{Cui2020} is a re-simulation of a mass-complete sample of 324 galaxy clusters from the MultiDark Planck 2 (MDPL2) $N$-body simulation~\citep[][]{Klypin2016}\footnote{\url{https://www.cosmosim.org/cms/simulations/mdpl2}} with $1 {\rm Gpc}/h$ simulation box size  \citep[see][for the benefit of this particular setup]{Zhang2022, deAndres2022NatAs, deAndres2023}. These clusters are identified with the \textsc{rockstar}~\citep{Rockstar} halo finder. A large high-resolution region $\sim 15 {\rm Mpc}/h$ in radius \citep[see][for the studies of filaments around clusters thanks to this large region]{Rost2021, Kuchner2020, Kuchner2021, Kuchner2022} at $z=0$ surrounding the central cluster is traced back to a very high redshift $z=120$ and the initial conditions (ICs) are generated by the parallel \textsc{GINNUNGAGAP} code\footnote{\url{https://github.com/ginnungagapgroup/ginnungagap}} including gas particles in the high-resolution region by splitting the dark matter particles and using low-resolution particles to sample the tidal fields outside of the high-resolution region in multiple layers.  From these ICs, three hydrodynamic simulation codes are used to evolve the clusters: \GM\ \citep{Sembolini2013}; \GX\ \citep{Rasia2015}; \GIZ\ \citep{Dave2016,Dave2019,Cui2022}, all of which are updated versions of \textsc{Gadget2} \citep{Springel2005}, a widely used N-body/SPH simulation code.  The three codes include models for baryonic processes such as gas cooling to form stars and  the enrichment of gas by supernovae. \GX\ and \GIZ also include black hole models and prescriptions for feedback from active galactic nuclei.  We refer readers \citet{Cui2022} for more detailed comparisons.

In this work, we use only the simulations from the \GX\ run, which agrees best with observed gas properties \citep{Li2021, Li2023}. 

The mass distribution of the main halos in The300 as a function of redshift is shown in Fig.~\ref{fig:mass_distribution}.  At $z = 0$, the distribution corresponds to a mass cut at $M \gtrsim 10^{15}  M_{\odot}$.  At higher redshifts, the mass distribution of the main halos becomes broader and shifts to lower values. Note that at high redshift, the main halo in a protocluster may constitute only a small fraction of the total mass of the structure \citep{Muldrew:2015}.  Furthermore, it is worth noting that although the sample is selected to be mass-complete at $z=0$, halos in The300 are not always the most massive halos at high redshift \citep[see][for the mass completeness at different redshifts in these simulations]{Cui2018,Cui2022EPJWC}. 

\subsection{SZ and lensing maps}
\label{sec:sz_lensing_maps}

To characterize the SZ signals from protoclusters we produce simulated Compton $y$ maps at several snapshots from The300 simulations. 
 Compton $y$ is related to the line-of-sight integral of the electron gas pressure, $P_e$, via
\begin{equation}
    y = \frac{\sigma_T}{m_e c^2} \int dl \,P_e(l),
\end{equation}
where $l$ is the line-of-sight distance.  We use $Y$ to represent the integral of $y$ over some circular aperture centered on the cluster:
\begin{equation}
\label{eq:bigY}
    Y = 2\pi \int_{0}^{R_{\rm max}} r\,dr\,y(r),
\end{equation}
where $r$ is the projected distance from the cluster center and $R_{\rm max}$ is the size of the aperture.  Following common convention, we set $R_{\rm max} = R_{200c}$.  The observed, frequency-dependent intensity change in the CMB maps due to the SZ effect is then proportional to Compton $y$ \citep{Carlstrom:2002}.  We will work entirely in Compton $y$ in this analysis, rather than frequency-dependent intensity. In principle, there may be some advantages to working with the raw intensity maps, particularly with regard to suppressing contamination from foregrounds \citep[e.g.][]{Chiang:2020}.  However, our choice greatly simplifies the analysis, and we expect it to have little impact on the forecast signal to noise, which is the main focus of this work.  For the protoclusters that we focus on, their masses are sufficiently small that the relativistic thermal SZ effect is expected to be negligible.  The kinematic SZ effect can also be ignored given its small amplitude relative to the thermal SZ.

In detail, these maps, which cover almost the full high-resolution region of the simulation (24 comoving ${\rm Mpc}/h$ on a side), are generated by the \textsc{PyMSZ} package\footnote{\url{https://github.com/weiguangcui/pymsz}} (see \citealt{Cui2018} for a more in depth description of the process of generating the mock $y$-maps). The angular resolution of the mock maps is initially fixed to 3$"$, i.e. much higher resolution than expected for the future CMB surveys of interest; we discuss the degradation of the resolution due to the telescope beam in \S\ref{sec:mock_obs}. Because the angular scale is fixed, the mock maps have different numbers of pixels and pixel sizes to cover the 24 ${\rm Mpc}/h$ on a side region at the different redshifts. 
The line-of-sight depth used when computing $y$ is also 24  comoving ${\rm Mpc}/h$; we discuss our inclusion of noise from foreground structure at larger distances in \S\ref{sec:mock_obs}. The mock maps are always centred on the centre of the whole simulation box, not at the most massive halos, which can be very far away from the simulation centre at high redshift. Though at low redshift, this region is much larger than the halo's virial radius, this guarantees that the whole protocluster region at high redshift is included.

\begin{figure*}
    \centering
\includegraphics[width=0.95\textwidth, keepaspectratio]{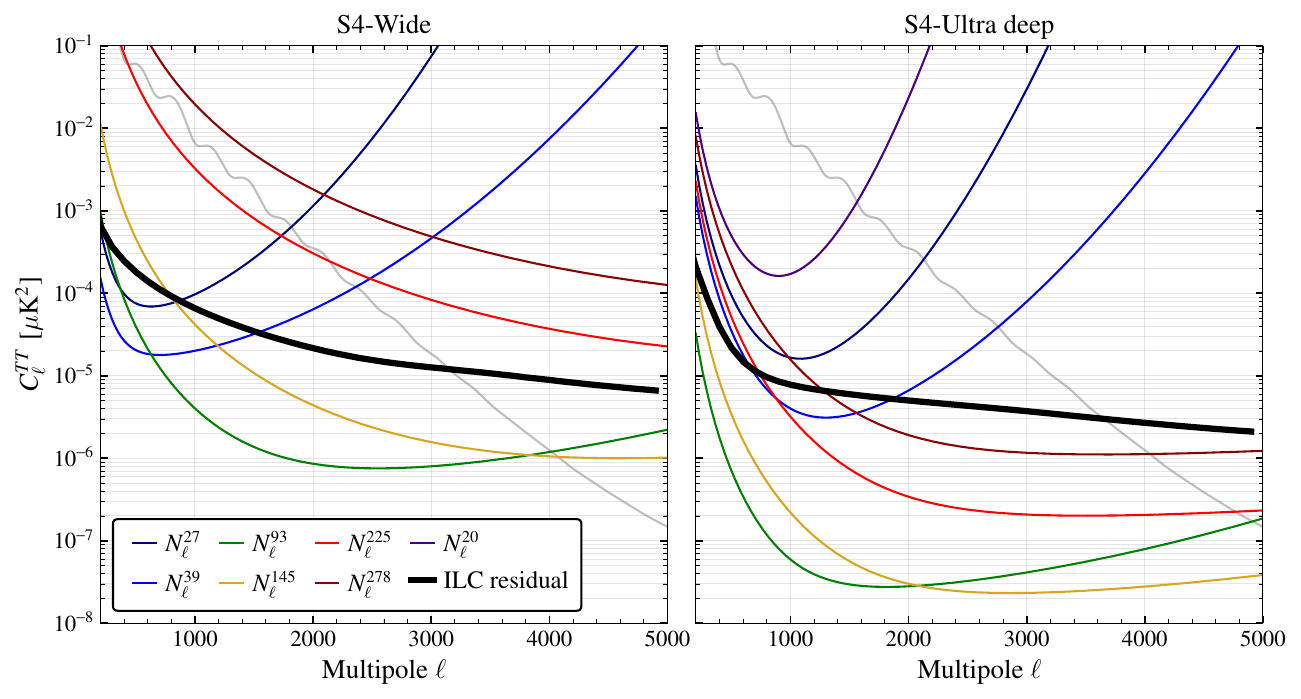}
    \caption{Power spectra of the residual noise and foregrounds after applying the minimum variance ILC weights to data from the \sfourwide{} (left panel) and \sfourdeep{} (right panel) surveys are shown using the thick black  curves. The thin coloured curves correspond to experimental noise in individual frequency bands. The lensed CMB power spectrum $C_{\ell}^{TT}$ is shown as the gray solid curves for reference.}
\label{fig:cmbs4_noise_ilc_curves}
\end{figure*}

\begin{figure}
    \centering
    \includegraphics[width=0.45\textwidth, keepaspectratio]{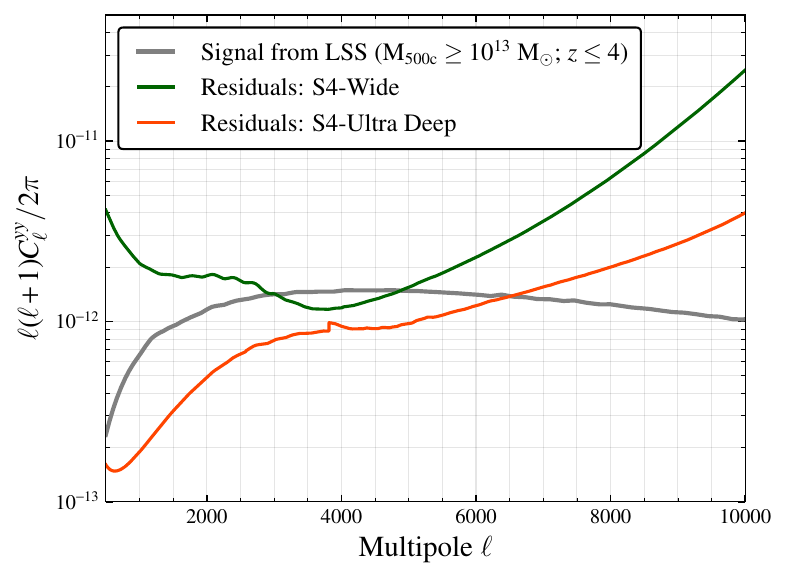}
    \caption{The total noise and foreground residuals expected in the MV Compton-$y$ map for \sfourwide{} (green) and \sfourdeep{} (red) surveys. The gray line indicates the Compton-$y$ power spectrum from large scale structure, which acts as an additional source of noise for the protocluster measurements. }
    \label{fig:compton_y_noise}
\end{figure}

\begin{figure}
    \centering
    \includegraphics[width=0.45\textwidth, keepaspectratio]{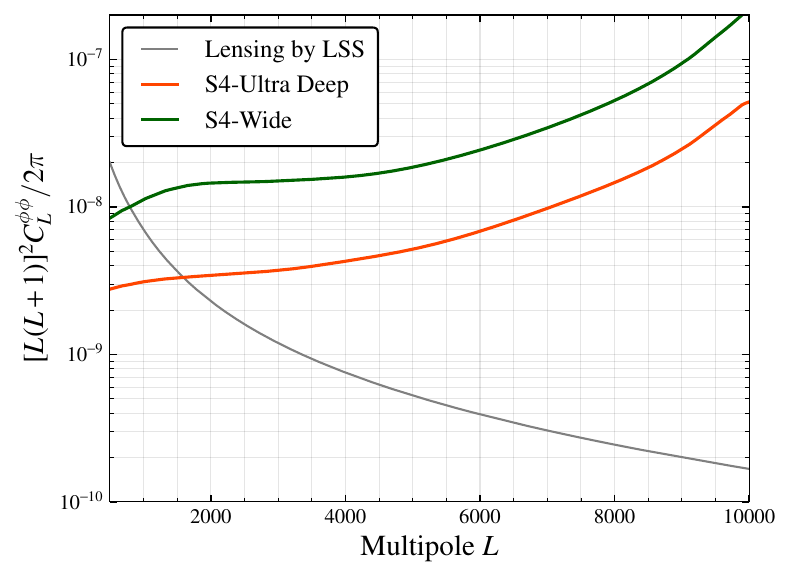}
    \caption{CMB lensing noise power spectra considered in this work.  The power spectra here are given in terms of the lensing potential $\phi$ rather than the convergence $\kappa$.  The green (red) curves represent the noise and foreground contributions for the \sfourwide{} (\sfourdeep{}) surveys, while the grey curve represents the lensing power spectrum from late-time large scale structure, which constitutes a source of noise for measurement of the protocluster lensing signal.}
    \label{fig:lensing_noise}
\end{figure}

The lensing maps are produced in a similar fashion to the SZ maps.  Under the Born approximation, the lensing deflection can be computed by integrating along the paths of undeflected photons.  In this case, the lensing convergence is related to the surface density in the plane of the lens, $\Sigma$, via
\begin{equation}
    \kappa = \frac{\Sigma}{\Sigma_{\rm crit}},
\end{equation}
where $\Sigma_{\rm crit}$ is the critical surface density given by
\begin{equation}
\Sigma_{\rm crit} = \frac{c^2 D_s}{4 \pi G D_l D_{ls}},
\end{equation}
where $D_s$ is the angular diameter distance to the source (in this case the last scattering surface, where the CMB originates), $D_l$ is the angular diameter distance to the lens, and $D_{ls}$ is the angular diameter distance between them.  We compute the surface density by integrating the simulation particle data along a randomly chosen line of sight. 

Note that $\kappa$ is not directly observable; rather, one measures the CMB temperature and polarization signals across the sky, and uses the statistical properties of these fluctuations to estimate $\kappa$ \citep[e.g.][]{hu02}.  Here, for simplicity we will work entirely with $\kappa$ rather than temperature or polarization maps.  

The SZ and lensing maps are generated at several redshifts: $z \in [3.5,  3,  2.5,  2.0, 0.9, 0.02]$ spanning the most interesting regime for detecting and characterizing protoclusters.

\subsection{Mock observations}
\label{sec:mock_obs}

In order to make forecasts for the ability of future CMB surveys to constrain the properties of protoclusters, we simulate the impact of noise, astrophysical foregrounds, and the telescope beam on the SZ and lensing observations.   We consider both the wide (\sfourwide) and deep (\sfourdeep) surveys from the CMB-S4 experiment \citep{cmbs4-sb1, cmbs4collab19}.

In addition to noise, CMB foregrounds such as the cosmic infrared background and radio sources present a significant challenge for estimation of Compton $y$ and CMB temperature/polarization (i.e. the relevant quantities for measuring CMB lensing).  To reduce the impact of foregrounds, one must rely on some form of component separation.  For our forecasts, 
we consider the use of internal linear combination (ILC) methods to estimate $y$ and the CMB temperature signal \citep[e.g][]{Tegmark:2003, Eriksen:2004}.  Such methods form a linear combination of the observations at different frequencies to estimate the component of choice.    Here, we choose weights in the linear combination such that there is unit response to the component of interest, i.e. CMB temperature (for the lensing forecasts) or $y$ (for the SZ forecasts) and the variance is minimized.  Note that for the SZ forecasts, the primary CMB fluctuations constitute a noise source.  We refer to this set of ILC weights as minimum variance or (MV).  This procedure accounts for additional variance introduced by foregrounds, but does not account for possible biases due to foregrounds, as we discuss below.  

We model the foregrounds using measurements made by SPT \citealt{reichardt21}). 
We include signals from primary CMB, kinematic SZ, radio galaxies and dusty star forming galaxies.  The noise is decomposed into atmospheric and instrumental parts.   The resultant temperature power spectra are shown in Fig.~\ref{fig:cmbs4_noise_ilc_curves}.
For more details about our noise and foreground modeling, we refer readers to \citet{raghunathan22, raghunathan22b}.  In Fig.~\ref{fig:compton_y_noise}, we show the combined noise and foreground power spectra in the Compton-$y$ map for \sfourwide{} in green and \sfourdeep{} in red. 

Since  Compton-$y$ is sensitive to the integrated pressure along the line of sight, large scale structure (LSS) will also contribute to the total variance when studying the signals from protoclusters.  The grey curve in the figure represents the  calculation of the power spectrum of Compton $y$ from LSS including halos greater than $M^{10}\,M_{\odot}$.\footnote{Our calculation of the LSS contribution includes only the one-halo term, which is expected to dominate across the scales of interest \citep{Komatsu:2002}.}  The total noise power used in our analysis is given by the sum of the LSS and residual noise and foreground contributions.    In principle, the LSS contribution to the noise could be reduced by identifying and masking halos with large SZ signals that are unassociated with protoclusters; by including the contributions from all halos above a low threshold we are therefore being conservative.   Note that the noise spectra shown in Fig.~\ref{fig:compton_y_noise} and Fig.~\ref{fig:lensing_noise} do not include the impact of the instrumental beam, which will suppress the noise at high $\ell$; we discuss the inclusion of the beam below.

Our estimate of the CMB lensing noise assumes that quadratic estimator (QE) methods \citep{hu02, okamoto03, hu07} are used to estimate the CMB lensing signal. 
The QE extracts the lensing signal using the lensing-induced correlations between different multipoles of the temperature and polarization fields (which are expected to be zero for a Gaussian and isotropic primordial CMB in the absence of lensing).  Chance correlations between different multipoles of the fluctuating CMB field, as well as instrumental noise, then contribute to the total CMB lensing noise.  We estimate the noise power of the CMB lensing maps following \citet{hu02}.

The presence of non-Gaussian foregrounds can also introduce such mode coupling, leading to biases in the reconstructed lensing signal.  
Particularly worrying is the SZ signal from the (proto)clusters themselves, which can bias the lensing estimates for low-redshift clusters \citep{Baxter:2015,Baxter:2018}.  
However, this bias is likely small for protoclusters given the expected $M^{5/3}$ scaling of the SZ signal.  
 Additional potential sources of bias include emission from galaxies residing within the protoclusters.  In general, recently developed methods may be able to mitigate some of these foreground biases with negligible impact to the signal-to-noise \citep{madhavacheril18, raghunathan19b, raghunathan19c, Hadzhiyska:2019,Horowitz:2019, levy23}. 
  Our CMB lensing noise estimate is derived assuming a combination of temperature and polarization measurements, and since the foreground signals are largely unpolarized, the lensing measurements based on polarization data are expected to be immune to foreground biases.  For these reasons, we will ignore potential biases in the lensing signal due to foregrounds, postponing a detailed study to future work.  Our final estimate of the CMB lensing noise power spectrum is shown in Fig.~\ref{fig:lensing_noise}.

Analogously to the SZ forecasts, we must also include the impact of CMB lensing fluctuations due to large scale structure, which constitute a noise source for our lensing forecasts.  We compute this contribution using \texttt{CAMB}\footnote{\url{camb.info}} \citep{Lewis:1999bs,Howlett:2012mh}.  This contribution is shown as the grey curve  in Fig.~\ref{fig:lensing_noise}.

Throughout this analysis, we assume that all sources of noise for our forecasts are statistically isotropic on the sky.  This is an excellent approximation for the extragalactic foregrounds.  Moreover, across each survey region (\sfourwide{} and \sfourdeep{}), CMB-S4 instrumental noise is expected to be close to uniform.  Galactic foregrounds, on the other hand, may introduce anisotropy into the noise.  We note, though, that our forecasts assume conservative values of sky coverage to account for removing regions of high galactic emission near the galactic plane.  Outside of these regions, the contributions of galactic foregrounds to the total foreground power are subdominant.

We generate Gaussian realizations of the SZ and lensing noise power spectra at the same pixel scale as the simulated SZ and lensing maps.   While the true noise is unlikely to be purely Gaussian,\footnote{For instance, the ``noise'' contributed by LSS must be non-Gaussian at some level.} this simple approach should be sufficient for the purposes of our forecasts.  To account for the impact of the telescope beam on the measurements, we apply Gaussian smoothing to the signal and noise maps\footnote{In reality, instrumental noise is not impacted by the telescope beam.  However,  in a real-space analysis like ours, one would typically smooth the noisy CMB lensing maps with an effective beam to suppress small-scale noise.  Our application of the beam to the noise maps reflects this procedure. } with $\theta_{\rm FWHM} = 1'$, close to the expected effective resolution of the CMB lensing maps from CMB-S4.  The smoothed signal and noise maps are then summed  to obtain the final mock-observed protocluster maps.  For essentially all of the protoclusters that we consider, the assumed beam size is smaller than the protocluster virial radii.  Consequently, we do not expect higher resolution observations to yield much improvement in signal-to-noise per protocluster (although higher resolution observations might be useful for, e.g., distinguishing features in protocluster SZ profiles).

\subsection{SZ and lensing profile measurements}
\label{sec_sz_lensing_profile_measurements}

A powerful way to characterize the lensing and SZ signals from protoclusters will be to measure their radial profiles.  We compute such profiles for each of the 324 simulated clusters by measuring the average of the lensing and SZ values in annuli centered on the protocluster, where the center is defined as the location of the maximum in the SZ or lensing maps.  At the resolutions accessible to future wide-field CMB experiments, the distinction between these two choices is unimportant.
The annuli used to measure the profiles are defined in  projected radius.  We use five bins from a minimum scale of $R_{\rm min} = 0$ to a maximum scale of $R_{\rm max} = 4\,{\rm Mpc}$ for both the SZE and lensing signals.  Note that in a real data analysis, using bins of projected radius (as opposed to angular bins) would require knowledge of the protocluster redshift, which could be obtained using e.g. spectroscopic followup.    
The CMB lensing and SZ measurements have noise power on large scales.  
These large-scale noise fluctuations will contribute to the covariance of the real-space profile measurements, increasing the diagonals and leading to large off-diagonals.  This is not a problem in principle for an analysis that includes the full covariance, but it does have the downside of making plots of the profile measurements much harder to interpret.  To reduce this effect, we subtract off a measurement of the local background from each of the protocluster profiles. This background estimate is derived from an annulus of width $0.8\,{\rm Mpc}$, with minimum radius equal to the largest radius profile measurement. 

In order to make forecasts for detectability of the SZ and lensing signals with CMB-S4, we compute the noise contribution to the profiles using the mock noise maps described in \S\ref{sec:simulations} using the same binning choices and background annulus subtraction as for the cluster measurements.  Since the noise is uncorrelated with the clusters and is additive, our estimated noise covariance for the profile of a single cluster is given by the covariance of the noise profiles. 

As noted previously, we also consider a stacking analysis in which we assume that protocluster positions have been determined via some other survey (e.g. a galaxy survey) and the signal-to-noise of the SZ and CMB lensing measurements is then enhanced by averaging across multiple protoclusters (which may or may not be individually detected in the CMB survey).    We form stacked profiles by averaging the individual protocluster profiles across all of the simulated protoclusters.   Two sources contribute variance to the stacked profile measurements: noise/foreground fluctuations in the SZ or CMB lensing maps, and intrinsic variation in the protocluster profiles themselves, both of which are included in our analysis.  The total covariance of the stacked profile measurements is computed from the sum of the covariances of the noiseless protocluster profiles and the noise-only covariance.  We compute the covariance of the stacked profile measurements using all clusters in The300, and then re-scale the errorbars to the expected number of observed clusters (see \S\ref{sec:cluster_number}).

\begin{figure*}
    \centering
    \includegraphics[width=\textwidth]{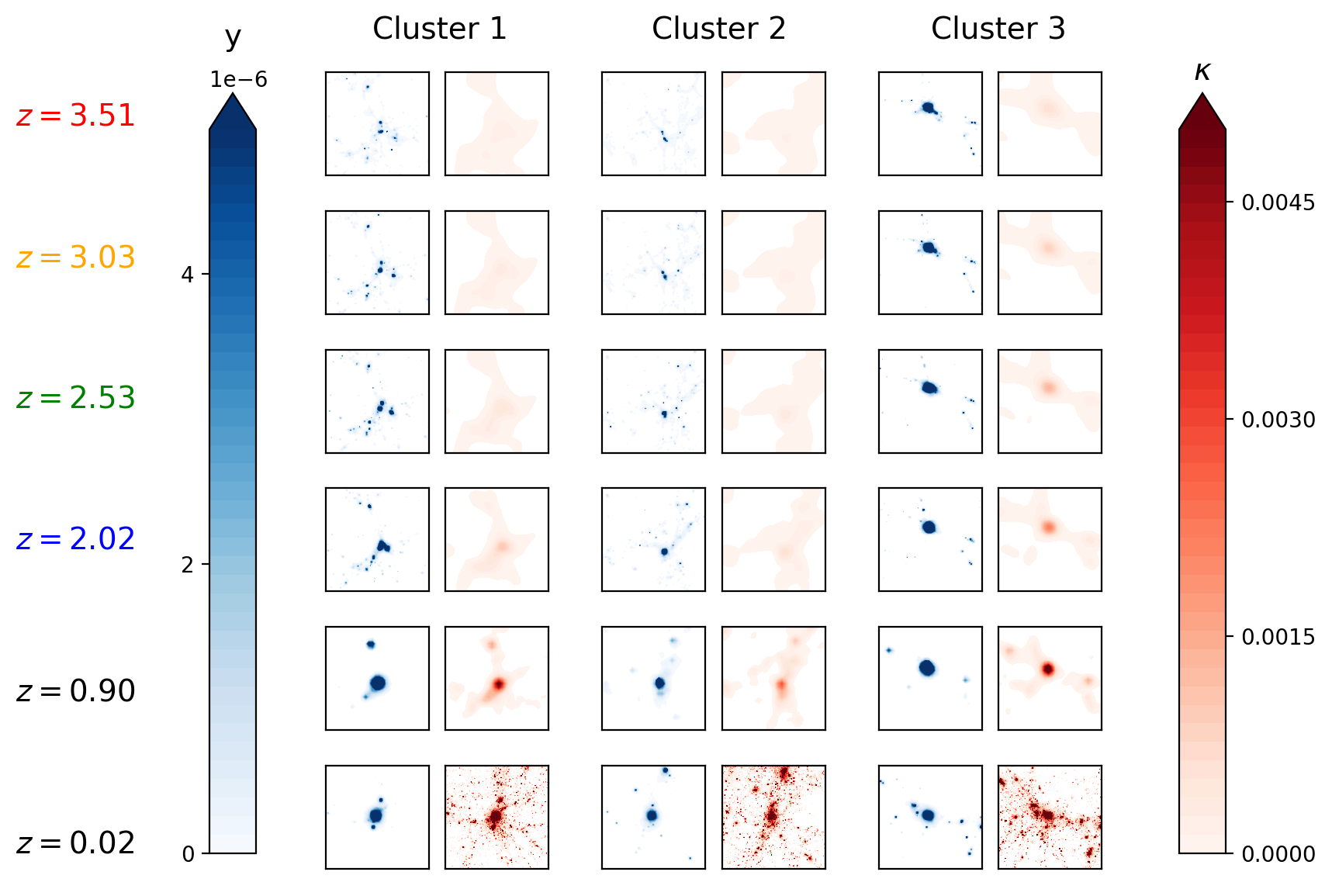}
    \caption{The evolution of three example protoclusters, as seen via the SZ effect ($y$, blue) and CMB lensing ($\kappa$, red).  Each panel shows a region 24 comoving $h^{-1} {\rm Mpc}$ on a side, and different rows correspond to different redshifts (indicated at left).  The color scales are clipped at high $y$ and $\kappa$.  
    }
    \label{fig:cluster_examples}
\end{figure*}

By chance, there is one very-high mass cluster in our sample that is essentially fully virialized at high redshift, leading to it having large SZ and lensing signals (cluster 3 in Fig.~\ref{fig:cluster_examples}).  This cluster contributes significantly to the estimated variance of the stacked profile measurements.  We therefore remove this cluster when computing averages across the full cluster sample. If necessary, a similar approach could also be taken in a real data analysis, with any anomalously high SNR clusters analyzed separately. 

\begin{figure*}
\centering
    \centering
    \subfloat[\centering Sunyaev Zel'dovich effect]{{\includegraphics[width=9cm]{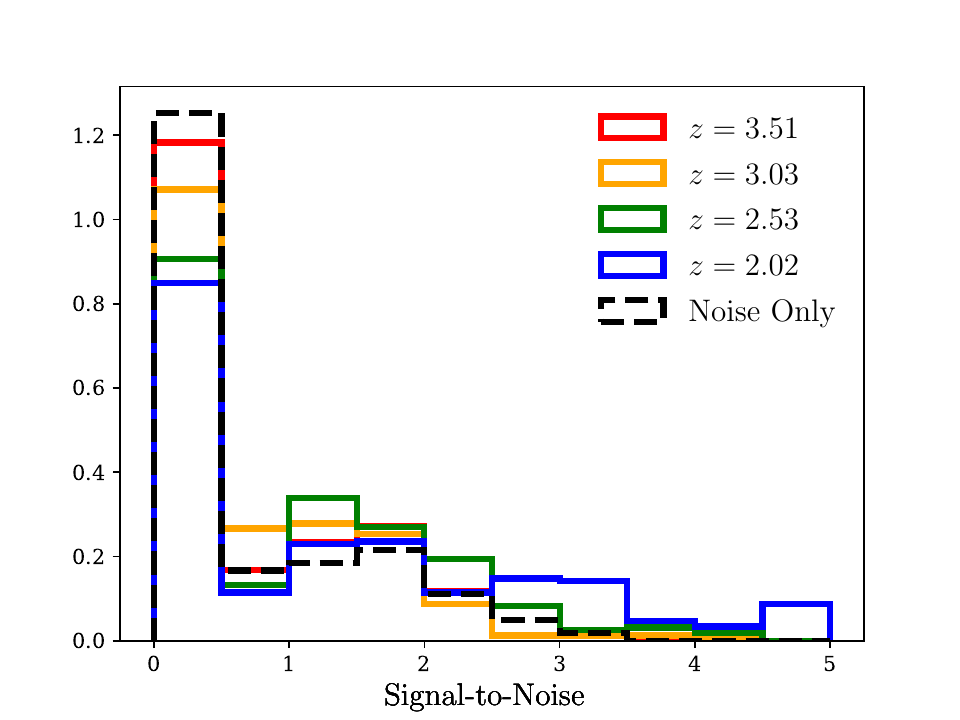} }}
    \subfloat[\centering CMB lensing]{{\includegraphics[width=9cm]{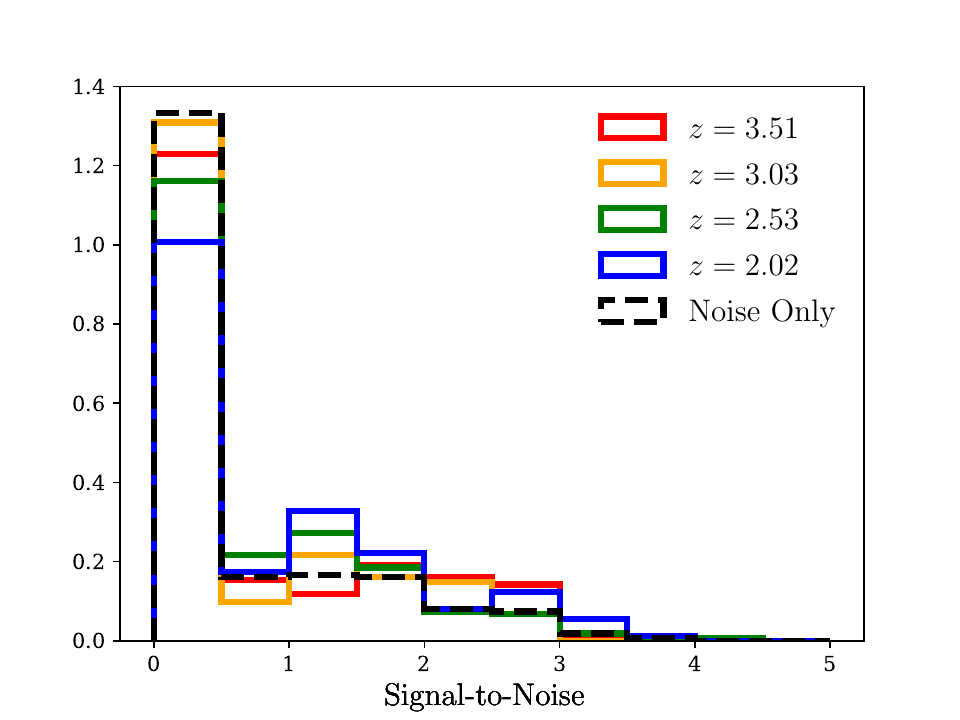} }}
    \caption{The signal-to-noise distribution of the SZ (left) and lensing (right) detections for individual simulated clusters (colored solid curves).  For reference, the black dashed curve represents the calculated signal-to-noise distribution when there is no cluster signal (i.e. noise only).  Values greater than zero are expected in this case due to random fluctuations.  Beyond $z \sim 2.5$, some protoclusters will be detected at high significance via their SZ signatures.  On the other hand, essentially no protoclusters can be detected beyond $z \sim 2$ via their CMB lensing signals. }
    \label{fig:snr_distribution}
\end{figure*}

\subsection{Expected number of protoclusters}
\label{sec:cluster_number}

The uncertainties on the stacked profile measurements depend on the number of protoclusters that we average over. Since we define protoclusters as the progenitors of a selection of redshift $z = 0$ clusters, the number of such objects per comoving volume remains fixed.  The number of protoclusters in the volume probed by CMB-S4 observations is then given by the number of clusters in The300 multiplied by the ratio of the comoving volume probed by CMB-S4 to the volume simulated (1 $({\rm Gpc}/h)^3$).  We adopt sky coverage of $f_{\rm sky} = 0.5$ for \sfourwide{} and $f_{\rm sky} = 0.03$ for \sfourdeep{} \citep{cmbs4collab19}.  We consider redshift bins of width $\Delta z = 0.1$ centered on the redshift values given above.  This bin width is small enough that the evolution of protoclusters across these slices is not dramatic, and large enough that there is a significant number of protoclusters in each redshift bin.  We find that the number of (proto)clusters in each redshift slice is then $N_c \approx [2600, 2700, 2600, 2500]$ for \sfourwide{}, and $N_c \approx [160, 160, 160, 150]$ for \sfourdeep{}.  Of course, these numbers will be different for different choices of $\Delta z$; also, assigning protoclusters to redshift bins will require obtaining redshifts, which may be obtained via spectroscopic follow up.

Note that this simple approach to estimating the number of protoclusters in the stack assumes that \text{all} protoclusters meeting our definition (i.e. that they give rise to clusters with $M \gtrsim 10^{15}\,M_{\odot}$ at $z = 0$) within the volume probed by the CMB-S4 observations are detected.  This assumption is likely to be unrealistic in several respects.  First, the threshold of $10^{15}\,M_{\odot}/h$ is largely arbitrary, set by choices made in the design of The300 simulations.  Still, this threshold is a reasonable and conservative choice in the sense that it corresponds to a very high mass cluster.  It is likely that future galaxy-based protocluster surveys will be able to probe the progenitors of less massive clusters, given that current surveys can identify protoclusters expected to be progenitors of clusters with $M \gtrsim 10^{14}\,M_{\odot}$  at $z = 0$ \citep{Toshikawa:2018}.  Second, any real detection of protoclusters will depend on their properties at the redshift they are observed, not on their properties at $z=0$.  For instance, a cluster with mass below our threshold at $z = 0$ might have evolved from a very massive protocluster that happened to have a sustained period of no accretion.  While we include some of the effects of different evolutionary histories (i.e. by averaging across an ensemble of progenitor objects), our treatment is not fully realistic because it ultimately relies on a cluster selection at $z = 0$.  
Finally, obtaining CMB observations of protoclusters detected in a galaxy survey will require overlapping galaxy and CMB survey footprints.   Fortunately, future galaxy surveys including the Vera C. Rubin Observatory Legacy Survey of Space and Time \citep[LSST;][]{LSST}, {\it Euclid} \citep{Euclid} --- will have significant overlap with CMB-S4, and have the potential to detect protoclusters \citep{Brinch:2023, Araya-Araya:2021}.  
Since the focus of this work is on using CMB surveys to detect and characterize protoclusters, we postpone a more detailed treatment of protocluster detection with galaxy surveys to future work.   Given some estimated number of protoclusters detected by a galaxy survey, one can simply scale our estimated errorbars by $\sqrt{324/N_c}$, where $N_c$ is the desired number of protoclusters.

\section{Results}
\label{sec:results}

\begin{figure}
    \centering
    \includegraphics[scale = 0.45]{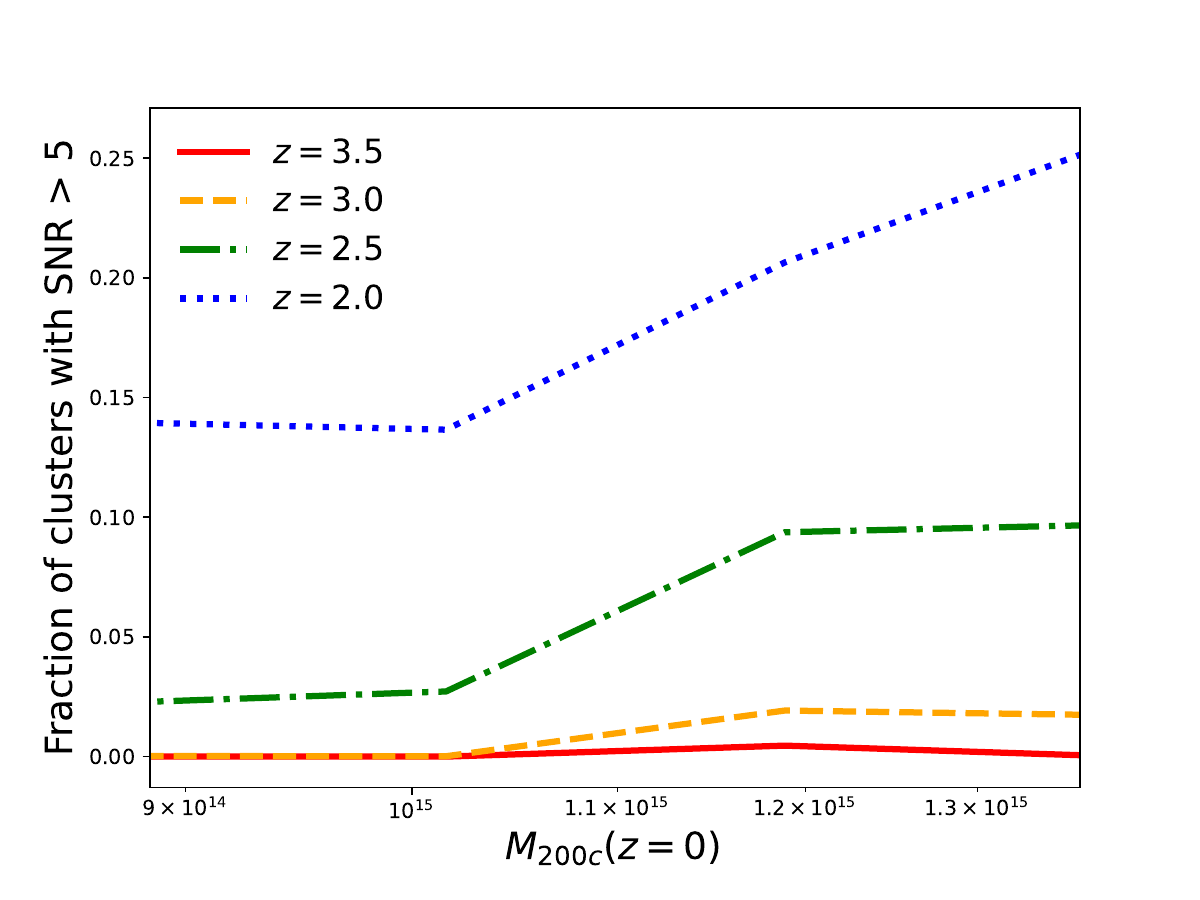}
    \caption{The fraction of protoclusters expected to have an SZ signal with ${\rm SNR} > 5$ at the indicated redshifts as a function of halo mass at $z = 0$.   
}
    \label{fig:frac_detected}
\end{figure}

\begin{figure*}
\centering
    \centering
    \subfloat[\centering Sunyaev Zel'dovich effect]{{\includegraphics[width=9cm]{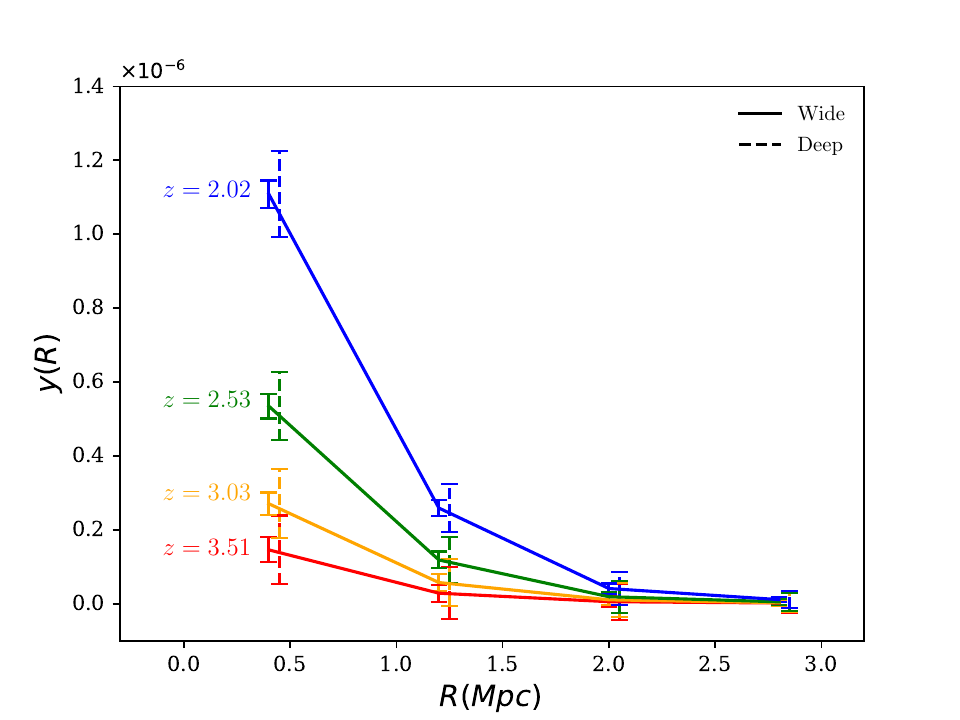} }}
    \subfloat[\centering CMB lensing]{{\includegraphics[width=9cm]{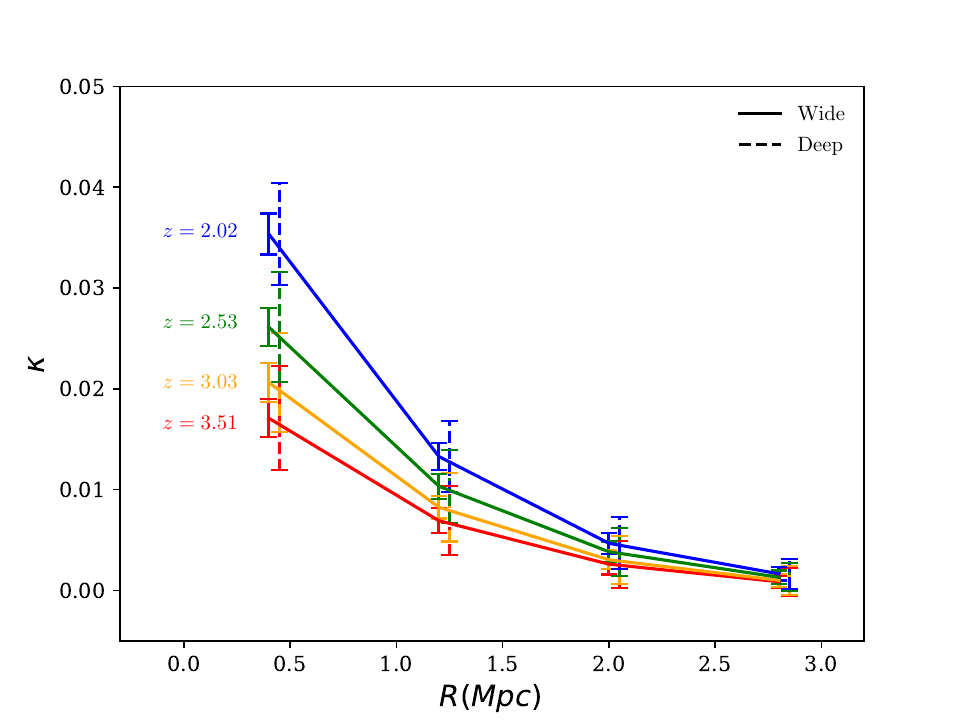} }}   
    \caption{Stacked SZ (left) and lensing (right) profile measurements.  We show errorbars corresponding to both \sfourwide{} (dotted) and \sfourdeep{} (solid); the errorbars for \sfourdeep{} are larger than for \sfourwide{} because of the smaller volume probed. }
    \label{fig:stacked_profiles}
\end{figure*}

We show the redshift evolution of the SZ and CMB lensing convergence maps in the absence of noise (but including the instrumental beam) for three different clusters in Fig.~\ref{fig:cluster_examples}.  Clusters 1 and 2 (columns one through four) are fairly typical, while cluster 3 (columns five and six) is the cluster with largest amplitude SZ signal of all The300 clusters.  At high redshift, the SZ and lensing signals are weak because halos within the protocluster have not assembled very much mass or gas.  By redshift of $z \sim 1$, the three structures have become approximately virialized galaxy clusters.   One can also see the filamentary structure surrounding the clusters, particularly visible in the lensing maps at low redshift.  

\subsection{Signal-to-noise of individual cluster measurements}

To assess the ability of future CMB surveys to detect protoclusters and to measure their individual properties, we compute the signal-to-noise of the mock SZ and lensing profile measurements for each cluster. We define the signal-to-noise of the $i$th cluster as 
\begin{equation}
{\rm SNR}_i \equiv \sqrt{d_i^{t} \mathbf{C} \vec{d}_i - N_d},
\end{equation}
where $\vec{d}_i$ is the profile measurement for the $i$th  cluster and $\mathbf{C}$ is the covariance of  the measurement, including both noise and intrinsic scatter), and $N_d$ is the number of elements in $\vec{d}$.
We subtract $N_d$ to prevent the estimated SNR from growing arbitrarily large as the length of the data vector increases, even in the absence of signal.  Note that we do not account for a ``look elsewhere'' effect associated with attempting to detect protoclusters at various positions across the sky; rather, the computed values correspond to the signal-to-noise at the true cluster location.   Note also that the we do not assume a profile shape when computing the signal-to-noise; the signal-to-noise could in principle be enhanced with the use of a matched filter.

Fig.~\ref{fig:snr_distribution} shows the distribution of cluster signal-to-noise values for the SZ (left) and lensing (right) measurements for \sfourdeep{}  We have excluded the highest signal-to-noise cluster (cluster 3 in Fig.~\ref{fig:cluster_examples}) when making this figure.  For comparison, we also show the signal-to-noise distribution when the profiles are computed using only the noise maps (black dashed curves).  For both the SZ and lensing measurements, it is clear that detecting protoclusters and characterizing their individual properties will be difficult, even with \sfourdeep{} observations: most protoclusters will not be detected with signal-to-noise greater than one.  For the SZ measurements,  which generally have higher signal-to-noise than the lensing measurements, there is a clear trend in the SNR distribution towards higher values as one goes to lower redshift, reflecting the build up of gas pressure in the main halos of the protoclusters over time.  Note that the evolution with redshift of the SZ signal-to-noise ratios is faster than for the lensing signal, since $Y \sim M^{5/3}$ while $\kappa \sim M$.  The steep dependence of the SZ signal with halo mass results from its dependence on both the electron number density, $n_e$, and  temperature,  $T$, which both increase with increasing halo mass ($Y \sim n_e T$, $n_e \sim M$

Although most protoclusters will not be detectable at high significance with \sfourdeep{}, some structures with large SZ signals may be detected: In Fig.~\ref{fig:snr_distribution}, a small but non-zero fraction of protoclusters will be detected with signal-to-noise greater than five.  To make this clearer, Fig.~\ref{fig:frac_detected} shows the fraction of protoclusters that will have high-significance SZ detections (${\rm SNR} > 5$) as a function of their mass at $z = 0$.  The significance of the SZ detection in this figure depends on both the noise level in the measurements (instrumental and foreground), and the evolutionary histories of the protoclusters.  Protoclusters that assemble a large fraction of mass at early times, for instance, will be easier to detect at high redshift.  We find that for clusters with $M \gtrsim 10^{15} \,M_{\odot}/h$ at $z =0$, the chances of obtaining a high-significance detection of the progenitor structure at $z \sim 2$ are roughly 20\%.  We find that essentially no protoclusters will have high-significance CMB lensing detections.  This finding appears to be roughly consistent with \citet{Madhavacheril:2017} and \citet{raghunathan22b}.  The typical mass of the main progenitor halos in our protocluster sample at $z \sim 2$ is $10^{13.6}\,M_{\odot}$; both \citet{Madhavacheril:2017} and \citet{raghunathan22b} predict that clusters with such masses will be below the detection threshold of CMB-S4.  Note, though, that both of those works assume models for the cluster SZ signal that are based on virialized low-redshift clusters.

\subsection{Stacked profile forecasts}

Since we have shown above that it will be difficult to detect and characterize individual protoclusters with future CMB observations, we now consider the prospects for a stacking analysis in which the protocluster measurements are averaged across many structures to improve signal-to-noise.  

In Fig.~\ref{fig:stacked_profiles}, we show the stacked lensing (right) and SZ (left) profiles measurements for the  protoclusters along with forecast errors for \sfourwide{} and \sfourdeep{}.  For \sfourwide{}, we forecast stacked SNR for the SZ measurements of $[16.5, 13, 7.2, 3.2]$ at redshifts of $[2.0, 2.5, 3.0, 3.5]$, respectively.  We find that the stacked measurements from \sfourwide{} will have somewhat higher precision than those of \sfourdeep{} given the larger number of clusters probed by \sfourwide{} (see discussion in \S\ref{sec:cluster_number}), although we emphasize that this result is sensitive the assumptions made in \S\ref{sec:cluster_number}.  For the stacked lensing measurements, the corresponding SNR values are $[11, 7.3, 5.7, 4.6]$.   The signal-to-noise values increase with decreasing redshift as the main halos of the protoclusters accrete more mass and gas.  We emphasize that it is these main halos to which the lensing and SZ signals are primarily sensitive: signals from mass outside these main halos makes a small contribution to the SZ and lensing signals, and in any case, this contribution is largely removed via the aperture subtraction process.  As explained previously, our forecasts assume statistically isotropic noise across the sky, and for this reason, we weight all clusters equally when computing the stacked profiles.  In principle, signals associated with the protoclusters themselves could lead to variation in the effective noise at the location of each protocluster.  However, since the individual protocluster signals are essentially undetectable, this is expected to be a small effect.

Interestingly, the lensing SNR in the highest redshift bin is actually higher than that for the SZ because of the rapid evolution of the SZ signal with halo mass noted previously.  The stacked lensing signal can also be detected to larger radii than the SZ signal for essentially the same reason: since the number density and temperature of the electron gas both decline with cluster-centric radius, the SZ signal declines faster with radius than the lensing signal.  The fact that the CMB lensing signal can be detected out to roughly 2.5 Mpc reflects the extended nature of protoclusters.    

\begin{figure}
    \centering
    \includegraphics[scale =0.42]{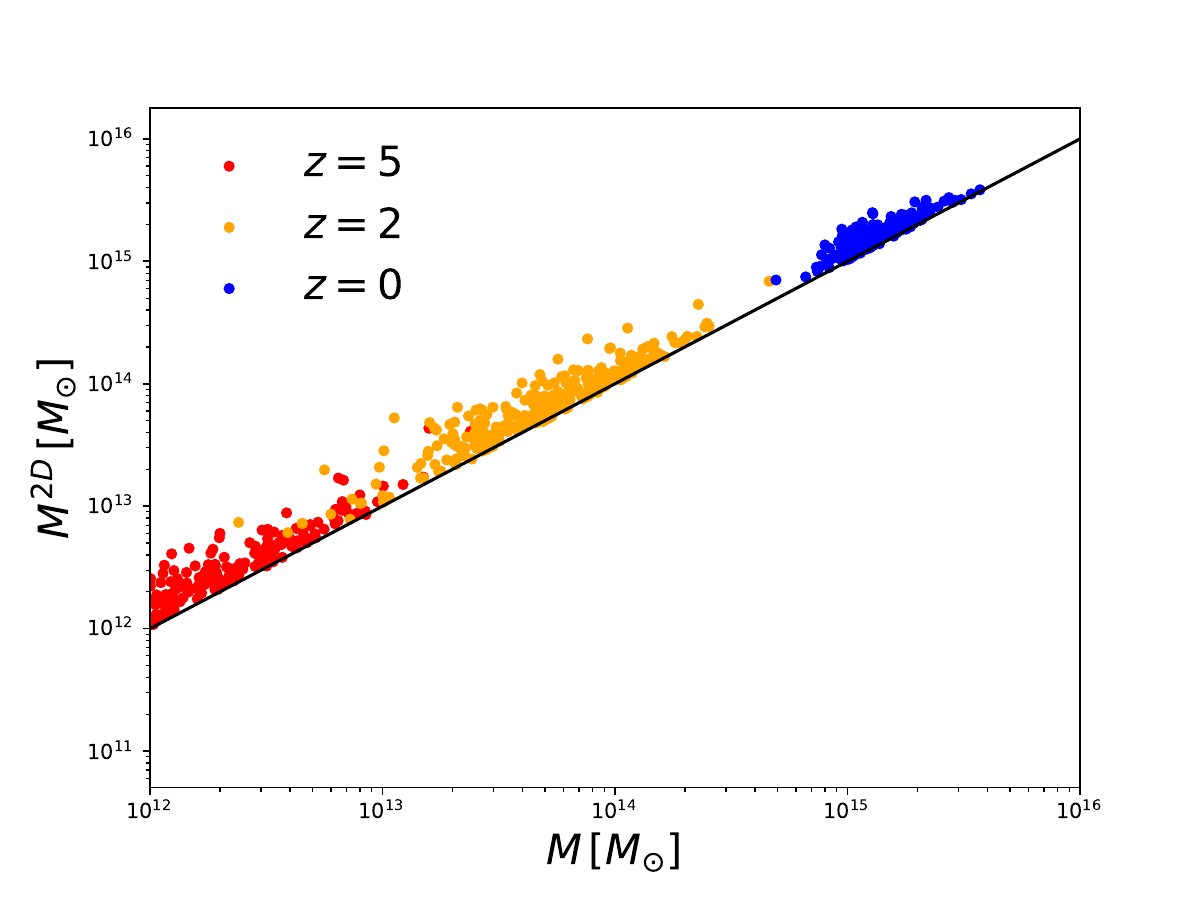}
    \caption{Relation between the true mass of the main halos within protoclusters, $M$, and the projected mass $M^{2D}$ that determines the gravitational lensing signal.  At high redshift, when protoclusters are still very extended, the scatter between 3D mass and projected mass becomes larger. }
    \label{fig:3d_vs_2d}
\end{figure}

\begin{figure*}
    \centering
\includegraphics[scale=0.4]{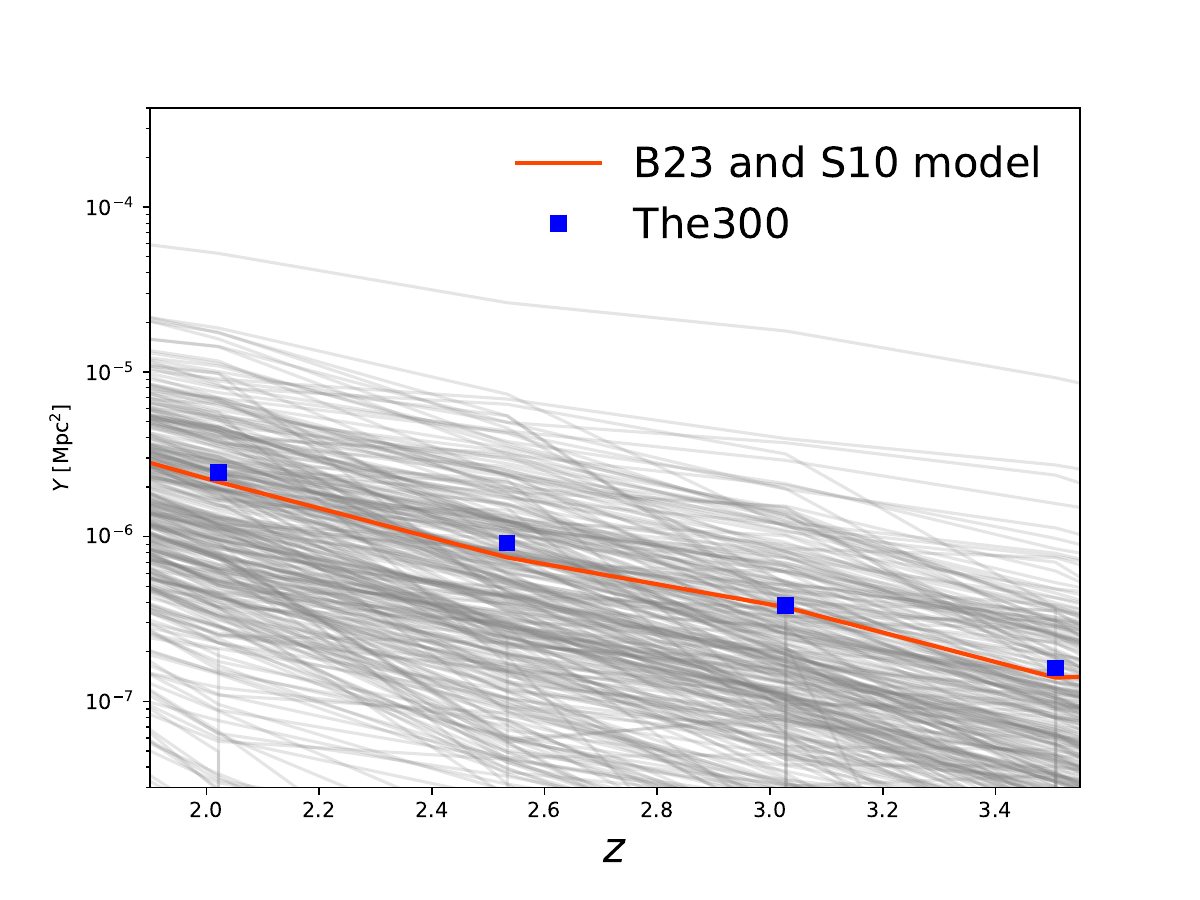}
\includegraphics[scale=0.4]{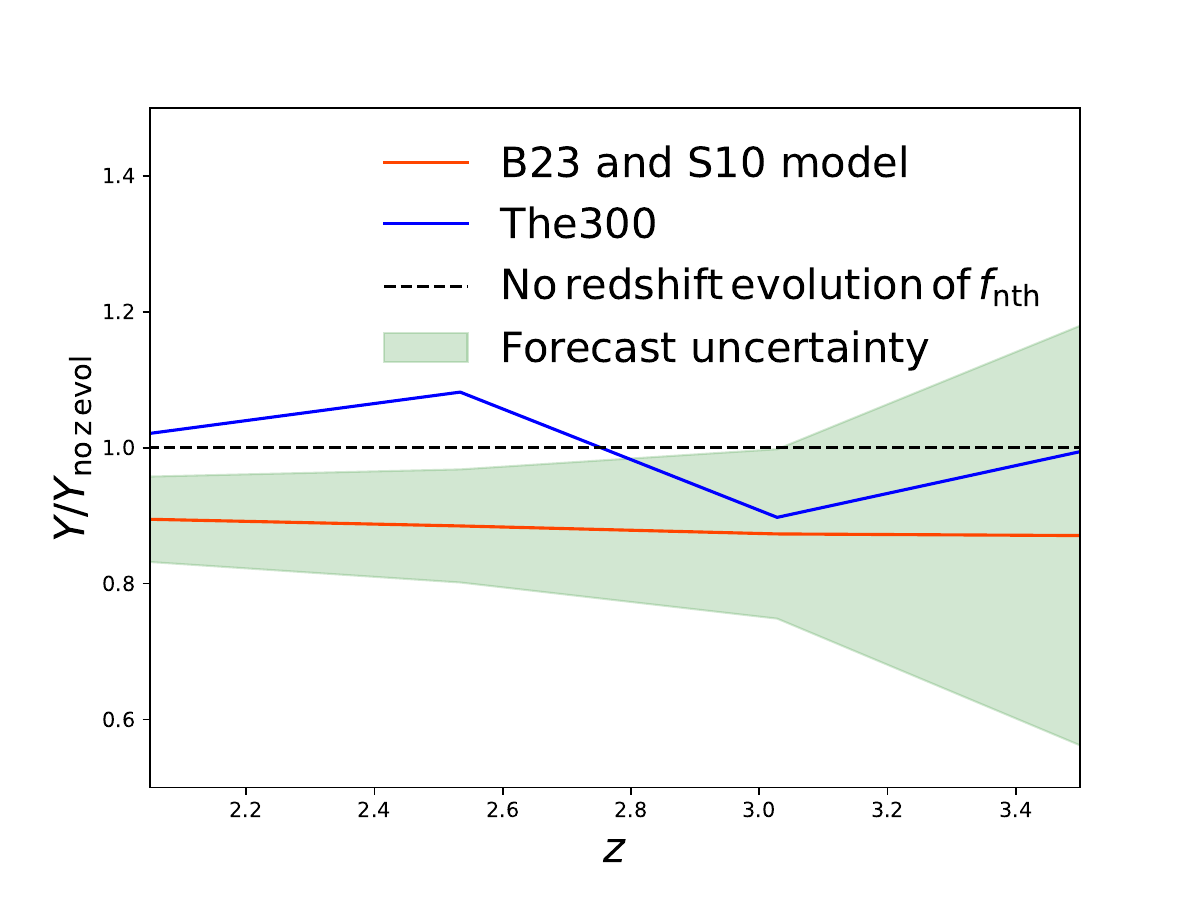}    
    \caption{Comparison of measured $Y$-$M$ relation to predictions based on the model of \citet{Baxter:2023} (B23), which includes the prescription for nonthermal pressure support from \citet{Shaw:2010} (S10) (orange curves in both panels).  Left panel shows the evolution of $Y$ with redshift for individual (main) halos from The300 (grey curves) and the average across all (main) halos in the The300 (blue points).  The right panel shows the same, but normalized to a model (black dashed curve) that fixes the nonthermal pressure fraction to its value at $z = 0$. The green band represents the forecast uncertainties on $Y$ from \sfourwide{}. }
    \label{fig:HSE}
\end{figure*}

Future SZ and CMB lensing protocluster profile measurements can be used to constrain the integrated $y$ signal, $Y$ (Eq.~\ref{eq:bigY}) and main halo mass, $M$, respectively.  We emphasize again that we are primarily concerned here with the properties of the main halo in the protocluster, which itself may be composed of many halos.  We compute $Y$ within an aperture of radius  $R_{\rm max} = R_{200c}$.  While $Y$ can be directly measured from the SZ observations, the lensing convergence measurements are sensitive to the line-of-sight projection of the protocluster mass, $M^{2D}$, rather than the true 3D mass:
\begin{equation}
  M^{2D} = 2\pi \int_{0}^{R_{\rm max}} \int dz \, r\,dr\,\rho(r,z),
\end{equation}
where $\rho(r,z)$ is the density at position $z$ and along the ray to the cluster, and $r$ is the radial separation from that ray.  As with $Y$, We choose an aperture size of $R_{\rm max} = R_{200c}$.
For low-redshift, virialized clusters, we expect $M^{2D}$ and $M$ to be tightly correlated, given that clusters are roughly spherically symmetric. Scatter in the relation between $M$ and $M^{2D}$ is expected due to e.g. triaxiality \citep{Rasia:2012}.  High-redshift protoclusters, on the other hand, may be far from spherical.  In Fig.~\ref{fig:3d_vs_2d} we show a comparison between  $M$ and $M^{2D}$ as a function of redshift for all simulated clusters.  The 2D mass is always larger than the 3D mass, since the volume probed is larger by definition.  As one goes to higher redshift, the 2D masses exhibit more scatter around $M$, reflecting the large asphericity of the protoclusters at high redshift.  The scatter around the 1-to-1 relation is roughly $\sigma(\log_{10}M^{2D}) = [0.05, 0.11, 0.12]$ at $z = [0, 2, 5]$, respectively.  In all cases, this scatter is lower than that due to noise in the CMB lensing  
measurements. In generating this figure, we exclude several clusters for which their evolution leads to them being near the edge of the high-resolution simulated region at high redshift.

\subsection{Constraining nonthermal pressure support and virialization}

Mergers and accretion events can drive significant bulk motion and turbulence in the cluster gas.  Bulk gas motions can in turn provide significant, nonthermal pressure support to the gas.  Unlike thermal motions, these bulk motions do not contribute to the thermal SZ signal.  Consequently, understanding the level of nonthermal pressure support in clusters, as well as its redshift evolution, is critical for many cosmological analyses of galaxy clusters \citep[e.g][]{Eckert:2019}.  Measuring the SZ signals of protoclusters and comparing to theoretical models therefore provides an important test of our understanding of cluster virialization, nonthermal pressure support, and the extent to which clusters depart from self-similar expectations at high redshift. 

Fig.~\ref{fig:HSE} shows the evolution of $Y$ as a function of redshift for the main progenitor halos of The300 clusters.  The grey curves represent individual halos, while the blue points represent the average across all halos.  It is clear that the evolution of $Y$ with redshift varies significantly between different clusters owing to their different evolutionary histories.  
The orange curve is the model which we describe below.

To model the evolution of $Y$ with redshift, we adopt the model for the $Y$-$M$ relation from \citet{Baxter:2023}, which in turn draws heavily from \citet{KomatsuSeljak2001, Komatsu:2002}; we refer readers to those works for more details of the modeling choices, and provide only a summary here.  The \citet{KomatsuSeljak2001, Komatsu:2002} model assumes that the mass distribution of the halo is described by a Navarro-Frenk-White profile \citep{NFW}, and that the halo gas is in hydrostatic equilibrium within this potential.  The amount of gas in the halo is set equal to the cosmic baryon fraction, and the gas density profile is assumed to follow the matter profile at large cluster-centric distances.  The \citet{Baxter:2023} model additionally includes a prescription for the decline in the halo pressure profile due to virial shocks at the halo splashback radius  \citep{Diemer:2014}, as predicted by \citet{Bertschinger:1985}.  This decline can have a significant impact on the $Y$-$M$ relation. 
 The \citet{Baxter:2023} model also includes a simple prescription for the two-halo term, which accounts for the contributions to $Y$ from nearby halos \citep[e.g.][]{Vikram:2017}.  Finally, the \citet{Baxter:2023} model incorporates a prescription for nonthermal pressure support which is in turn derived from \citet{Shaw:2010}.  Briefly, the fraction of the total pressure supplied by nonthermal sources is assumed to have a radial and redshift dependence of the form
\begin{equation}
    f_{\rm nth}(r,z) = \alpha_{\rm nt}(z)(r/R_{500c})^{0.8},
\end{equation}
where
\begin{eqnarray}
\alpha_{\rm nt}(z) &=& \alpha_0 f(z) \\
f(z) &=& {\rm min}\left[(1+z)^{\beta}, (f_{\rm max} - 1) {\rm tanh}(\beta z) + 1 \right] \\
f_{\rm max} &=& 4^{-n_{\rm nt}/\alpha_0},
\end{eqnarray}
with $\alpha_0 = 0.18$ and $n_{nt} = 0.8$.  This model was found by \citet{Shaw:2010} to provide a good match to the simulations of \citet{Nagai:2007}.  \citet{Shaw:2010} also found weak evidence for redshift evolution of $n_{\rm nt}$ as well as dependence of the redshift evolution of nonthermal pressure on the details of the baryonic modeling.

The orange curves in both panels of Fig.~\ref{fig:HSE} represent the calculation of $Y$ using the \citet{Baxter:2023} model with the \citet{Shaw:2010} prescription for nonthermal pressure support.  In calculating this model curve, we adopt the true mean mass of the main halos of the protocluster structures at each redshift.  We find that this model provides a remarkably accurate description of the average redshift evolution of $Y$ over a wide range of redshift (left panel of Fig.~\ref{fig:HSE}).  This is perhaps surprising given that protoclusters can significantly violate many of the assumptions of the \citet{Baxter:2023} model.  It appears, though, that while individual protoclusters may exhibit significant departures from e.g. a NFW density profile leading to significant scatter in their $Y$ evolution, these differences tend to wash out when averaged over many clusters.  In the right panel of Fig.~\ref{fig:HSE} we plot the redshift evolution of  our baseline model (again with the orange curve) relative to a model for which we fix $\alpha_{\rm nt}(z) = \alpha_0 f(0)$, i.e. for which there is no redshift evolution in the nonthermal pressure fraction.  Between redshifts of $z\sim 2$ to $z \sim 5$, redshift evolution of the nonthermal pressure fraction is expected to suppress the mean $Y$ by about 10\%.  This is because in the model of \citet{Shaw:2010} (see also \citealt{Stanek:2010}), the nonthermal pressure fraction increases with redshift (up to some maximum value), leading to a suppression of $Y$.  The green band in the right panel of Fig.~\ref{fig:HSE} shows the expected uncertainty on $Y$ given the stacked measurements from \sfourwide{}.  It appears that the precision offered by CMB-S4 will be sufficient to test a plausible range of predictions for the redshift evolution of nonthermal pressure support.  Note, though, that uncertainties on mass estimates for protoclusters may complicate such an analysis.  We postpone a detailed investigation of  this possibility to future work.  The blue curve in the right panel of Fig.~\ref{fig:HSE} represents the evolution of $Y$ as measured in The300.  There is significant scatter in these measurements, particularly at high redshift, due to intrinsic variation between the clusters (see also the left panel of the figure).  To within this scatter, The300 clusters agree reasonably well with the redshift evolution of $Y$ predicted by the \citet{Shaw:2010} model, although there may be a hint of slightly less redshift evolution of the nonthermal pressure support. 

\section{Summary}
\label{sec:discussion}

We have explored the ability of future CMB surveys like CMB-S4 to detect and characterize the properties of high-redshift protoclusters.  Relative to previous studies, a unique feature of our analysis is that we use realistic simulated protoclusters from The300 in our forecasts, rather than analytic models based on low-redshift clusters.  This is important because the assumptions of e.g. spherical symmetry and hydrostatic equilibrium can be grossly violated for protoclusters.  

We find that it will be challenging to detect the lensing and SZ signals from individual protoclusters, even with high-sensitivity data from \sfourdeep{}.  However, we do expect that of order 10\% of the progenitors of very massive clusters ($M \gtrsim 10^{15} M_{\odot}$) will have SZ signals detectable at high significance at $z = 2.5$ (see Fig.~\ref{fig:frac_detected}).  It is unlikely that any protoclusters at high redshift will be detected via their CMB lensing signals.

To date, most protoclusters have been detected via overdensities of high-redshift galaxies.  Assuming that surveys like {\it Euclid},  LSST and the Nancy Grace Roman Space telescope \citep{Spergel:2015} can identify samples of protoclusters via galaxy overdensities in the region surveyed by CMB-S4, it will be possible to constrain their average gas content and mass using CMB-S4 (Fig.~\ref{fig:stacked_profiles}).  Assuming all protoclusters that are progenitors of clusters at $z=0$ with $M \gtrsim 10^{15} \,M_{\odot}$ can be identified, we find that the CMB lensing signal, which is directly related to the cluster mass, can be measured with a signal-to-noise of roughly 5.7 at $z \sim 3$, while the SZ signal can be measured with signal-to-noise of 7.2 at this redshift.  Because of their non-spherical nature, protoclusters will exhibit significantly more scatter in their lensing (i.e. projected) masses than low-redshift virialized clusters (Fig.~\ref{fig:3d_vs_2d}).  However, this scatter will be small compared to the uncertainties from the CMB lensing mass estimates.

Finally, the measurements of the protocluster SZ signals with \sfourwide{}  can be used to test models for the redshift evolution of nonthermal pressure support within protoclusters (Fig.~\ref{fig:HSE}), providing insight into the processes by which galaxy clusters assemble and virialize.

We emphasize that several of the forecasts in this work are likely conservative, since our stacking analysis is predicated on being able to identify only those protoclusters which are the progenitors of clusters at $z = 0$ with $M \gtrsim 10^{15}\,M_{\odot}$.  It is likely that future wide-field galaxy surveys with significant overlap with \sfourwide{} will be able to detect less massive structures.  Our forecasts can be trivially scaled to account for any desired number of protoclusters; we postpone investigation of more realistic protocluster selections to future work.

\section{Data availability}

This manuscript was developed using data from The300 galaxy clusters sample. The data is available on request following the guideline of The300 collaboration (\url{https://www.the300-project.org}). The data used to make the figures shown in this work are available upon request.

\section*{Acknowledgments}
This work was partially supported by the Center for AstroPhysical Surveys (CAPS) at the National Center for Supercomputing Applications (NCSA), University of Illinois Urbana-Champaign. 

This work made use of the Illinois Campus Cluster, a computing resource that is operated by the Illinois Campus Cluster Program (ICCP) in conjunction with the National Center for Supercomputing Applications (NCSA) and which is supported by funds from the University of Illinois at Urbana-Champaign.
This work also used the computational and storage services associated with the Hoffman2 Shared Cluster provided by UCLA Institute for Digital Research and Education's Research Technology Group.

We thank Lukas Zalesky for fruitful discussions related to this work.

EB is partially supported by NSF Grant AST-2306165.

WC is supported by the STFC AGP Grant ST/V000594/1 and the Atracci\'{o}n de Talento Contract no. 2020-T1/TIC-19882 granted by the Comunidad de Madrid in Spain. He also thanks the Ministerio de Ciencia e Innovación (Spain) for financial support under Project grant PID2021-122603NB-C21 and ERC: HORIZON-TMA-MSCA-SE for supporting the LACEGAL-III project with grant number 101086388.

\bibliographystyle{aasjournal}
\bibliography{protoclusterbib}

\end{document}